# Giant magnetoimpedance: new electrochemical option to monitor surface effects?

Galina V. Kurlyandskaya[a], Vanessa Fal Miyar[b]

[a]Dept.of Electricity and Electronics, University of the Basque Country UPV-EHU, Apdo. 644, 48080, Bilbao, Spain,

[b]Dept. of Physics, University of Oviedo, Calvo Sotelo s/n, 33007, Oviedo, Spain

*Keywords: Magnetoimpedance; Biosensors; Electrochemical magnetoimpedance spectroscopy*

Tel. +34 94 601 3237; Fax: + 34 94 601 3071; E-mail: galina@uniovi.es

## 1. Introduction

In the last years, various analytical devices incorporating biological materials have been developed in order to satisfy increasing requirements for sensitivity and specificity (Turner, 2000). In 1998 the idea of using a magnetic field as a transducer for simultaneous detection of many molecular recognition events was reported and the success of the designed sensor based on magnetoresistance technology stimulated intensive research in the field of magnetic biosensors (Baselt et al., 1998). In the majority of magnetic biosensors, albeit not always (Ong, et al., 2001), magnetic particles play an essential role, being employed as biomolecular labels (Ferreira, et al., 2002; Miller et al., 2002; Lany, et al., 2005). Giant magnetoimpedance, MI, was proposed recently (Kurlyandskaya, et al., 2003), for the detection of magnetically labelled biomolecules. In the reported prototypes of MI biosensors, stable compositions without protective covering were used for sensitive elements (Kurlyandskaya et. al, 2003; 2005) as well as those using a protective layer (Bethke, et. al., 2003). The idea of a MI biosensor based on amorphous ribbon is attractive since this would allow construction of a cheap detector with the sensitive element designed as a disposable strip. In early studies, FeCoMoSiB ribbons were shown to be stable for certain conditions of MI biosensing. At the same time there are other materials which show a high MI effect, but suffer rapid degradation when subjected to direct contact with biofluids. Thorough testing of different compositions could provide answers to questions as to which materials, and in which detecting conditions, can be proposed as MI sensitive elements without a protective covering.

Electrochemical impedance spectroscopy, EIS, is a powerful method for monitoring a wide variety of chemical and physical properties in order to analyse resistive and capacitive responses to stimulation of the system by sinusoidal alternative current (Katz, et al., 2003). It has been used as a research tool in many areas such as coatings, corrosion phenomena, conducting polymers and biosensors. In this last case, EIS enables monitoring of biological reactions at the electrode surfaces (Guan et al., 2004).

The MI effect consists of the change in the impedance of a ferromagnetic conductor, Z, under application of the magnetic field when high frequency alternating current flows through it (Beach, et al., 1994). The main features of MI were explained by means of Maxwell equations solved for certain geometries and particular models (Landau, et al., 1975). High frequency impedance depends on classic skin penetration depth, $\delta$, i.e. $Z = Z(\delta)$:

$$\delta = (\rho/\pi\mu f)^{1/2}, \qquad (1)$$

where $\rho$ is the electrical resistivity, $\mu$ is transverse magnetic permeability, and f is the frequency of the ac current. In non-magnetic materials, skin depth can be controlled by changing the frequency. In magnetic materials, the application of a magnetic field changes the magnetic permeability, skin depth, and as a result, the total impedance. MI in materials with the appropriate anisotropy shows extraordinary sensitivity of the order of one hundred percent per Oersted. Therefore one can propose the magnetoimpedance effect as a new electrochemical option for biosensing in order to probe the electric features of surface-modified magnetic electrodes, in the expectation of very high sensitivity when the biological reactions, the material of the sensitive element, and the detection conditions are properly selected and synergetically adjusted.

This work is a part of the most recent development of a biosensor working in a principle of electrochemical magnetoimpedance spectroscopy. A prototype has been designed and tested using amorphous ribbon of specially selected composition as its sensitive element. The MI measurements were successfully completed both in a regime where chemical surface modification by biofluids and sensing were separated and in a regime where they were not separated (in a bath for fluids).

## 2. Materials and methods

Rapidly quenched amorphous ribbons of two compositions were selected for testing. The geometries and properties of all samples are collected in Table 1. The amorphous structure in as-quenched state, and after surface modifying treatments, was checked by x-ray diffraction using Cu-$K_\alpha$ radiation. The surface morphology before and after treatments was studied by optical microscopy using a Nikon L-UEPI microscope. Quasistatic longitudinal magnetization curves were measured for dry ultra sound cleaned samples by a conventional inductive technique at a frequency of 0,9 Hz.

Different fluids were used for surface modification of the ribbons: distilled water; phosphate buffer saline, sugar solution in the distilled water, PBS, physiological solution and human urine (Table 2). The pH of all solutions was checked at 22$^o$C, i.e. for the same temperature as that selected for all studies. Urine was collected from supposedly healthy male volunteers, filtered and always used at a temperature of 22 $^o$C. The first regime of the surface modification consisted in exposure for a controlled time, in on-edge position, with no stir of the solution. The mass of the samples was measured after cautious rinsing, ultra sound washing in distilled water during 2-4 minutes, and drying. The relative mass loss for initial mass $m_0$ was defined as follow: $\Delta m/m_0 = 100 \times (m_0 - m)/m_0$. The average mass loss per hour was recalculated on the basis of 500 hours exposure to each fluid.

The magnetoimpedance was measured by a standard four point technique with constant amplitude of the sinusoidal driving current $I_{rms} = 60$ mA, at frequencies of 0.5 to 10 MHz. Additional details on the method can be found elsewhere (Kurlyandskaya, et al., 2005). Electric contacts between the sample and Cu-leads of the imprinted circuit were prepared by Ag conductive paint. An external magnetic field, H, was applied in the plane of the ribbon parallel to the ribbon long side. The MI measurements were always carried out in decreasing (down branch) and increasing (up branch) fields, starting at the maximum positive field. The magnetoimpedance ratio, $\Delta Z/Z$ was defined as follows:

$$\frac{\Delta Z}{Z} = \frac{100(Z(H) - Z(H_{max}))}{Z(H_{max})}, \qquad (2)$$

where $H_{max} = \pm 73$ Oe. The maximum value of MI ratio, which appears in the field close to the anisotropy field, was denominated as $\Delta Z/Z_{max}$. It took about 25 minutes to measure each of the complete MI curves.

Due to the extreme magnetic softness of the samples (coercive fields, $H_c$, are less than the Earth field component), the MI curves were slightly biased by effective laboratory fields. This did not affect the results of the present research but should be mentioned in order to explain why MI curves show, in this particular case, 0,25 Oe displacement towards the negative fields (Fig. 1, 3, 4, 6). The lowest value of the MI ratio in the field close to 0,25 Oe was denominated as $\Delta Z/Z_{min}$. For convenience the field of - 0,25 Oe will be called the "close to zero field".

The ribbon sensitive element was installed in the printed circuit board, with or without a bath for fluids, for testing a whole device as a biosensor-prototype. A non-magnetic plastic bath with inner dimensions of 52×10×7 (mm) was installed when necessary in the central part of the board. The MI measurements were made under two regimes: X – "passive surface modification" where chemical surface modification by biofluids and sensing were separated; Y - in a bath for fluids, i.e. where chemical surface modification by biofluids was continuously monitored by MI data collection.

Surface modification in the areas of the electrical contacts could negatively affect the impedance measurements because of the contacts' lower quality. In order to avoid this problem the ends of the samples prepared for the MI measurements under regime X were covered by protective glue before treatment in the biofluids. This was subsequently removed by solvent, after the treatment. As a result the surface of the ribbon in the central part of 62 mm was modified but the 19 mm ends were kept un-affected, thus providing good electric contacts.

The measurements under regime Y, with a bath for fluids, required a strict maintenance protocol. Once the MI measurements had started the surface of the ribbon was always kept covered by fluid. In order to change the fluid, approximately 75% of the volume was first removed and then replaced with fresh fluid and this whole procedure was repeated 3 times. After each change, an interval of 10 minutes was allowed before starting the next appropriate MI curve measurement. The bath with Ortho-Phosphoric acid was refreshed every 1,5 hours; the urine bath was refreshed every 12 hours.

**3. Results and Discussion**

*3.1. Characterization of the as-quenched ribbons*

Co-based amorphous ribbons with close to zero negative magnetostriction are well-studied MI materials (Dmitrieva, et al., 1998). In the first MI biosensor prototype for magnetic label detection (Kurlyandskaya et. al, 2005), FeCoMoSiB ribbons were shown to be stable for sensing in phosphate buffer saline. But they also required additional relaxation heat treatment at 340 $^o$C. Any post–preparation treatment can be considered to be a technological disadvantage. Therefore as-quenched ribbons were selected for this research: $Fe_5Co_{70}Si_{15}B_{10}$, assumed to be non-stable, and $Fe_3Co_{67}Cr_3Si_{15}B_{12}$, assumed to be very stable in low and high pH fluids (Table1).

The quasistatic hysteresis loops and MI responses were measured under regime X for $Fe_5Co_{70}Si_{15}B_{10}$-II and $Fe_3Co_{67}Cr_3Si_{15}B_{12}$-II samples in as-quenched state (Fig. 1). Both ribbons are magnetically soft: $H_c$ of about 0,06 and 0,05 Oe were found for $Fe_5Co_{70}Si_{15}B_{10}$-II and $Fe_3Co_{67}Cr_3Si_{15}B_{12}$-II samples respectively. In both cases $\Delta Z/Z_{max}$ appears in the fields close to the anisotropy field of about 0,9 Oe. For all frequencies under consideration the MI effect is higher for $Fe_3Co_{67}Cr_3Si_{15}B_{12}$-II samples. For example, for f = 5 MHz the $\Delta Z/Z_{max} \approx 280$ % for $Fe_3Co_{67}Cr_3Si_{15}B_{12}$-II and $\Delta Z/Z_{max} \approx 160$ % for $Fe_5Co_{70}Si_{15}B_{10}$-II ribbons. Features of both the hysteresis loops (the slope of the M(H) in low fields before saturation) and MI curves (a two-peaked shape with a pronounced valley in the field close to zero) indicate that the effective magnetic anisotropy is not exactly longitudinal. Such behaviour is quite well-known. It is commonly accepted to ascribe it to non-uniform magnetic anisotropy when the surface anisotropy is different compared with that at the centre of the ribbon (M. Vazquez, et al., 1999).

Ribbons of both compositions show very small MI hysteresis and the presence of two different field intervals, where MI responses can be approximated by linear $\Delta Z/Z(H)$ dependence. The sensitivities of the MI responses were calculated for the work points corresponding to the central parts of these intervals (points A and B in the inset of the Figure 1d): for the point A H(A) = 1,2 ± 0,4 Oe the $\Delta Z/Z$ sensitivities were found to be 45 and 14 %/Oe for $Fe_3Co_{67}Cr_3Si_{15}B_{12}$-II and $Fe_5Co_{70}Si_{15}B_{10}$-II ribbons respectively; for the point B H(B) = 0,1 ± 0,04 Oe, the $\Delta Z/Z$ sensitivities were found to be 460 and 420 %/Oe for $Fe_3Co_{67}Cr_3Si_{15}B_{12}$-II and $Fe_5Co_{70}Si_{15}B_{10}$-II ribbons respectively. This means that although the total MI effect is smaller in $Fe_5Co_{70}Si_{15}B_{10}$ ribbons there is a field interval were it shows very high sensitivity which is almost comparable to that of the $Fe_3Co_{67}Cr_3Si_{15}B_{12}$ ribbons. In this interval the MI behaviour is assumed to be very much dependent on the features of the surface anisotropy and therefore MI monitoring could be a way to reveal reactions at the surface.

*3.2. Passive surface modification (regime X)*

For the study of passive surface modification by different fluids, wide $Fe_3Co_{67}Cr_3Si_{15}B_{12}$-I and $Fe_5Co_{70}Si_{15}B_{10}$-I ribbons were selected in order to provide better accuracy. The results are collected in Figure 2 and Table 2. For all fluids under consideration, with no exception, the average mass loss for $Fe_3Co_{67}Cr_3Si_{15}B_{12}$-I ribbons is at least one order of magnitude smaller than that one for $Fe_5Co_{70}Si_{15}B_{10}$-I samples. For Cr-containing samples, the mass variation in distilled water or physiological solution can be considered as negligible. For PBS the average slope corresponding to $Fe_3Co_{67}Cr_3Si_{15}B_{12}$-I ribbons is negative in the context of the levels of accuracy attainable in this experiment. Although more precise study is necessary in order to prove this assumption, it suggests that passive exposure to PBS resulted in the formation of a protective layer which prevented any loss of mass. Distilled water was the only solution which caused non-linear mass loss of the $Fe_5Co_{70}Si_{15}B_{10}$-I ribbons as a function of the exposure time.

Urine or sugar containing urine caused far greater loss of mass at a much faster rate compared with the slopes corresponding to distilled water, sugar solution in distilled water, PBS, or physiological solution. The x-ray study (inset of the Figure 2d shows such an example), confirms the amorphous state of the ribbons before and after passive surface modification in urine. The appearance of a wide peak at small angles, corresponding to non-crystalline reaction products on the surface, is also seen.

As a very preliminary hypothesis for further study, we note that a very slight difference in mass loss was observed when comparing the cases of urine and sugar containing urine (Figure 2d, Table 2). One of the arguments supporting the interest of this hypothesis is the change in visually observable optical properties: in the case of urine with sugar, the characteristic violet tones appeared for $Fe_3Co_{67}Cr_3Si_{15}B_{12}$ ribbon treated for a few hundred hours, but did not appear in the case of pure urine.

$Fe_3Co_{67}Cr_3Si_{15}B_{12}$ ribbon shows excellent stability under regime X of passive surface modification and can be proposed as a stable sensitive element for MI-prototypes based on magnetic label detection in distilled water, in sugar water solution, in physiological solution and, especially, in PBS. By contrast, $Fe_5Co_{70}Si_{15}B_{10}$-I ribbons do not show high stability and can, therefore, be proposed as candidates for MI-prototypes based on the electrochemical principle.

Both $Fe_3Co_{67}Cr_3Si_{15}B_{12}$-I and $Fe_5Co_{70}Si_{15}B_{10}$-I show rather fast, close to linear mass loss when subjected to passive surface modification in Ortho-Phosphoric acid. The rate of mass loss depends on the concentration and it was always higher for $Fe_5Co_{70}Si_{15}B_{10}$ ribbons. Figure 4a collects these data for wide samples (see also Table 2). For the next step a 7% concentration which provided reasonably fast, but controllable, mass loss was selected. Although a reasonable correlation of the mass loss rate for samples of the same composition but slightly different geometries was expected, the difference in the rate of mass loss was specifically checked for $Fe_3Co_{67}Cr_3Si_{15}B_{12}$-I and $Fe_5Co_{70}Si_{15}B_{10}$-I and –III samples. At first glance the narrow sample can be characterized by a higher mass loss rate (curve B), but it gets very close to that of the wide sample when normalized so as to take into account the difference of surface to volume ratio (curve B and curve C).

*3.3. Active surface modification by Ortho-Phosphoric acid combined with MI measurements (regime Y)*

$Fe_5Co_{70}Si_{15}B_{10}$-III ribbon was installed in a bath for fluids. The prototype was calibrated with no fluids in a bath for frequencies of 3, 4, and 5 MHz. Then MI responses were measured with distilled water and a frequency of 4 MHz, close to that providing the maximum $\Delta Z/Z$ ratio, was selected for measurements with $H_3PO_4$ 7% acid. Complete MI curves were measured periodically, keeping a time record throughout all measurements and, to begin with, a time count at the moment of the bath being filled with the acid. Some examples of MI curves are shown in the Figure 3. MI responses are clearly dependent on the time of the exposure to the $H_3PO_4$ 7% acid. In order to visualize this dependence, two characteristics of the $\Delta Z/Z$ curves were selected: $\Delta Z/Z_{max}$ and $\Delta Z/Z_{min}$. Figure 4c shows that MI maximum $\Delta Z/Z_{max}$ shows linear decay, but the $\Delta Z/Z_{min}$ parameter can be very satisfactorily adjusted by Lorentz fit. The difference between $\Delta Z/Z_{max}$ and $\Delta Z/Z_{min}$ is maximum for the initial state but it becomes less and less pronounced with increased exposure, i.e. not only absolute values of certain parameters of the MI curve but also its shape, depend on the exposure time. As had been expected, the clearest changes of shape appeared in an area corresponding to the responses in a small field – the central valley of the $\Delta Z/Z$ curve related to surface anisotropy in this particular case. It was possible to monitor the change of the surface properties of the $Fe_5Co_{70}Si_{15}B_{10}$ ribbon under the electrochemical regime by MI spectroscopy.

*3.4. MI measurements of the $Fe_5Co_{70}Si_{15}B_{10}$-II ribbons surface modified in urine by passive method (regime X)*

In order to provide both accurate MI measurements and observation of the surface morphology, $Fe_5Co_{70}Si_{15}B_{10}$-II ribbons were selected for regime X experiments with urine. First of all the ribbons were checked to have similar geometrical parameters, magnetic properties and MI responses in initial state. Than the central part of the samples (see section 2 for more details) was subjected to passive treatment in urine for 7, 22, 38, 45, 46, 120, 145, 210, or 460 hours and MI of each sample after cleaning, drying, and installation into the imprinted circuit board was measured in a frequency range of 0.5 to 10 MHz. Fig. 4 a shows the examples of MI responses for selected treatments. Again MI responses are clearly dependent on the length of the exposure to the urine and the central part of the curve in a small field is more subject to change as exposure increases.

Although general changes in the shape of the MI curve have similar tendencies, such as less and less pronounced differences between $\Delta Z/Z_{max}$ and $\Delta Z/Z_{min}$ with the increase in exposure, there is a specific feature which was not observed in the case of Ortho-Phosphoric acid. This is the non-monotonous behaviour of both $\Delta Z/Z_{max}$ and $\Delta Z/Z_{min}$: for low exposures, i.e. under approximately 45 hours they increase, and then show a fast decrease up to approximately 140 hours exposure, followed by a period of very slow decrease (Fig4 b,c). These features were observed for all frequencies above 1 MHz (Fig. 4b). This behaviour is in accordance with the proposed surface anisotropy model. Usually very thin surface layers, of the order of 1-2μm, of the as-quenched ribbons are the areas where both higher internal stresses and higher magnetic anisotropy deviations appear. Linear mass loss (Fig. 2d) is not affected by these particular anisotropy features but MI responses do have this sensitivity. For low exposures the mass loss corresponds to the removal of a highly stressed surface layer. It is somehow equivalent to surface relaxation, which increases magnetic softness and the total value of the MI ratio. When a surface layer is partially or totally removed, MI monitoring still has sufficient sensitivity to follow the surface modification process which most probably depends on the initial surface structure, the type of biofluid and on temperature.

The optical microscopy observations reveal the surface modification features. Figure 5 shows examples of the $Fe_5Co_{70}Si_{15}B_{10}$-II ribbon surfaces after passive modification in urine. It can be seen that observations of the free side (as opposed to the roller side) are much more informative. The first obvious changes in the surface state were observed after 6 hours exposure (not shown here). After 45 hours the whole surface area was significantly modified, usually shown by vertically oriented affected areas of different grey tones. In the central part of Figure 6c, a non-typical black area was specially recorded

– these areas of profound surface change appear with high frequency after 120 hours treatment and transform into a principal structure after about 210 hours.

Although the results of the measurements for urine modified samples in regime X look satisfactory, the data shown in Figure 4c are not precise. The high level of experimental error is due to the many parameters involved and which need to be controlled: initial properties of the ribbons, identity of the surface modification treatments, contacts quality, etc. Therefore, in order to demonstrate the capability of MI spectroscopy as an electrochemical method that could be adapted for biosensing, the surface modification combined with direct MI measurement was made for human urine.

*3.5. Active surface modification by urine combined with MI measurements (regime Y)*

MI measurements of the $Fe_5Co_{70}Si_{15}B_{10}$-II ribbons combined with surface modification in urine had shown higher accuracy compared with measurements in regime X and similar tendencies: non-monotonous behaviour of both $\Delta Z/Z_{max}$ and $\Delta Z/Z_{min}$ with very little increase in separation after the maximum achieved near 40 hours (Figure 6). Both $\Delta Z/Z_{max}(t)$ and $\Delta Z/Z_{min}(t)$ can be reasonably fitted by Gauss or Lorentz curves. The $\Delta Z/Z_{min}(t)$ curve shows close to linear dependence and higher sensitivity in the period of 15-35 hours (part A in Figure 6b).

The existence of the peak of the $\Delta Z/Z_{max}(t)$ and $\Delta Z/Z_{min}(t)$ curves in case of the urine and its absence in case of the $H_3PO_4$ study can be explained by very different mass loss rates caused by these fluids. It may be that, due to the very high rate in the Ortho-Phosphoric acid case, the thin surface layer responsible for the $\Delta Z/Z$ increase at the beginning of the treatment had simply been removed before the first MI measurement was completed.

MI Monitoring of the surface treatment during surface modification is assumed not to be a passive treatment, i.e. regimes X and Y can be different. The surface structure observed by optical microscopy indicates a possible difference (Fig.5 g, h). The regime Y surface structure corresponding to 101 hours of treatment does not have an analogue for regime X of 0 to 515 hours structures. Regime Y structure is very uniform over the whole surface area, it has clear white almost non-affected zones and very black areas similar to those which appear under regime I after 120 hours of treatment. These black and white zones have a slight tendency to be oriented less than $45^o$ to the ribbon axis. Dark field observation (Figure 6 h) does not reveal special contrast in the borders between the white and black areas, a fact which increases the probability of these parts of the structure being rather flat.

The hysteresis loops of the ribbon before and after treatment are very different. After treatment under regime Y, a two phased M(H) loop appears. The behaviour in a small field is very close to that which was observed in the initial state – magnetization of soft magnetic material in a direction close to the easy magnetisation axis. In a field of about 0,28 Oe this behaviour changes completely and probably can be best described as magnetization of soft magnetic material in a direction close to the hard magnetization axis. One can assume a correlation between the appearance of the black areas observed by optical microscopy (which could have a transverse magnetization component) and the change observed in the hysteresis loop.

Another argument in support of the proposed surface anisotropy model comes from the estimates of a skin penetration depth for $Fe_5Co_{70}Si_{15}B_{10}$ ribbon in accordance with eq. (1). In a state of high permeability (very small external magnetic fields) the skin depth is close to half the thickness of the sample, i.e. the surface starts to play an important role and this is why the method is sensitive to the surface state. At the same time, a few percent mass loss in case of the human urine does not seem to cause a critical change in the sample thickness/skin depth ratio and therefore it seems that the observed effects can be ascribed to the change of the surface anisotropy rather than to the relative change of thickness with respect to skin depth.

$Fe_5Co_{70}Si_{15}B_{10}$ as-quenched amorphous ribbons can be considered as a promising material for robust devises based on electrochemical magnetoimpedance spectroscopy and therefore they can act as the interface between biological materials and electronics.

## 4. Conclusions

The MI effect was studied in as-quenched $Fe_5Co_{70}Si_{15}B_{10}$ and $Fe_3Co_{67}Cr_3Si_{15}B_{12}$ amorphous ribbons for a driving current of 60 mA in the interval of 0.5 to 10 MHz. Although the total effect was always smaller in $Fe_5Co_{70}Si_{15}B_{10}$ ribbons, it was found that in a low field, the MI sensitivity was almost comparable for $Fe_5Co_{70}Si_{15}B_{10}$ and $Fe_3Co_{67}Cr_3Si_{15}B_{12}$ ribbons.

The $Fe_3Co_{67}Cr_3Si_{15}B_{12}$ ribbon shows excellent stability in a regime of passive surface modification and can be proposed as a stable sensitive element for MI-prototypes based on magnetic label detection in distilled water, physiological solution and, especially, PBS without deposition of the protective layer. By contrast, $Fe_5Co_{70}Si_{15}B_{10}$ ribbon can be proposed as a candidate for MI-prototypes based on the electrochemical principle.

It was shown that the magnetoimpedance effect can be successfully used as a new electrochemical option to probe the electric features of the magnetic electrode surfaces modified by Optho-Phosphoric acid and urine when the biofluid, the material of the sensitive element, and the detection conditions are properly selected and synergetically adjusted.

**Acknowledgements**

This work was supported by a "Ramon y Cajal" grant of Spanish MEC and UPV-EHU and in part by an ACTIMAT grant of The Basque Country Government. Some of the measurements were made during experimental sections of the course "Biosensors: their implementation and impact on the quality of life" and we should, therefore, like to thank the PhD students of the Basque Country University, Estibaliz Asua and Javier Rodriguez, for their technical assistance. Special thanks to V.I. Levit, A.P. Potapov, A. Garcia-Arribas, J.M. Barandiaran, J.A. Garcia, and K.J. Kalim for their support.
measurements were made during experimental sections of the course "Biosensors: their implementation and impact on the quality of life" and we should, therefore, like to thank the PhD students of the Basque Country University, Estibaliz Asua and Javier Rodriguez, for their technical assistance. Special thanks to V.I. Levit, A.P. Potapov, A. Garcia-Arribas, J.M. Barandiaran, J.A. Garcia, and K.J. Kalim for their support.

Table 1. Composition, geometry and properties of amorphous ribbon samples: $\lambda_s$ – magnetostiction coefficient, $\rho$ - resistivity, $M_s$ – saturation magnetization.

| Composition | Geomery | $\lambda_s \times 10^7$ | $\rho$ ($\mu\Omega$cm) | $4\pi M_s$ (kGs) |
|---|---|---|---|---|
| $Fe_5Co_{70}Si_{15}B_{10}$ | I -   $0,025 \times 2 \times 100$ mm$^3$<br>II -  $0,019 \times 1,2 \times 100$ mm$^3$<br>III – $0,018 \times 0,9 \times 100$ mm$^3$ | $\sim -1$ | 134 | 6,7 |
| $Fe_3Co_{67}Cr_3Si_{15}B_{12}$ | I -   $0,025 \times 2 \times 100$ mm$^3$<br>II -  $0,019 \times 1,2 \times 100$ mm$^3$ | $\sim -1$ | 148 | 4,7 |

Table 2. Description of the test solutions and results of the passive surface modification treatments.

| Test solution | Sample type | Fit | Mass loss after 500 hours (%) | Average mass loss (%/h) | Visually estimated colour |
|---|---|---|---|---|---|
| Distilled water (pH 7,0) | $Fe_5Co_{70}Si_{15}B_{10}$ | Polynomial | 3,80 | $7,6\times10^{-3}$ | Very dark yellow |
| | $Fe_3Co_{67}Si_{15}B_{12}$ | Linear | | $\times10^{-4}$ | Non very uniform light change of initial grey to mate with yellow |
| Distilled water + sugar (0,05 g/ml) | $Fe_5Co_{70}Si_{15}B_{10}$ | Linear | 1,31 | $2,6\times10^{-3}$ | Dark yellow with tendency to mate |
| | $Fe_3Co_{67}Si_{15}B_{12}$ | Linear | 0,12 | $3,3\times10^{-4}$ | Non very uniform light change of initial grey to mate |
| Physiological solution (pH 7,4) | $Fe_5Co_{70}Si_{15}B_{10}$ | Almost linear | 2,45 | $5,0\times10^{-3}$ | Relatively uniform change to semi transparent yellow |
| | $Fe_3Co_{67}Si_{15}B_{12}$ | Linear | 0,13 | $2,6\times10^{-4}$ | Light change to dark grey with yellow |
| Phosphate buffered saline (pH 7,3) | $Fe_5Co_{70}Si_{15}B_{10}$ | linear | 0,9 | $1,1\times10^{-3}$ | Very light change to semitransparent grey with moonlight |
| | $Fe_3Co_{67}Si_{15}B_{12}$ | linear | 0,02 | $-1,7\times10^{-4}$ | Almost the same |
| Urine (pH 6) | $Fe_5Co_{70}Si_{15}B_{10}$ | linear | 16,3 | $3,3\times10^{-2}$ | Non uniform mate black |
| | $Fe_3Co_{67}Si_{15}B_{12}$ | linear | 0,25 | $5,0\times10^{-4}$ | More dark than initial grey with mixed blue and yellow |
| Urine (pH 6)+ sugar (0,05 g/ml) | $Fe_5Co_{70}Si_{15}B_{10}$ | linear | 14,1 | $2,8\times10^{-2}$ | Black; semi mate |
| | $Fe_3Co_{67}Si_{15}B_{12}$ | linear | 0,45 | $9,0\times10^{-4}$ | More dark than initial grey with blue and violet |
| 7% Ortho-Phosphoric acid (pH 0,92) | $Fe_5Co_{70}Si_{15}B_{10}$ | linear | - | 2,5 | Metallic grey; mate |
| | $Fe_3Co_{67}Si_{15}B_{12}$ | linear | - | 5,3 | Metallic grey; mate |

Figure 1.

a) Inductive hysteresis loops.
b-c) MI responses of the amorphous ribbons for selected frequencies at $I_{rms}$= 60 mA.
d) Comparison of the MI behaviour of Co-based ribbons for a frequency of 5MHz. Two possible work points of the sensor prototype are indicated: A – in a high field and B in a low field.

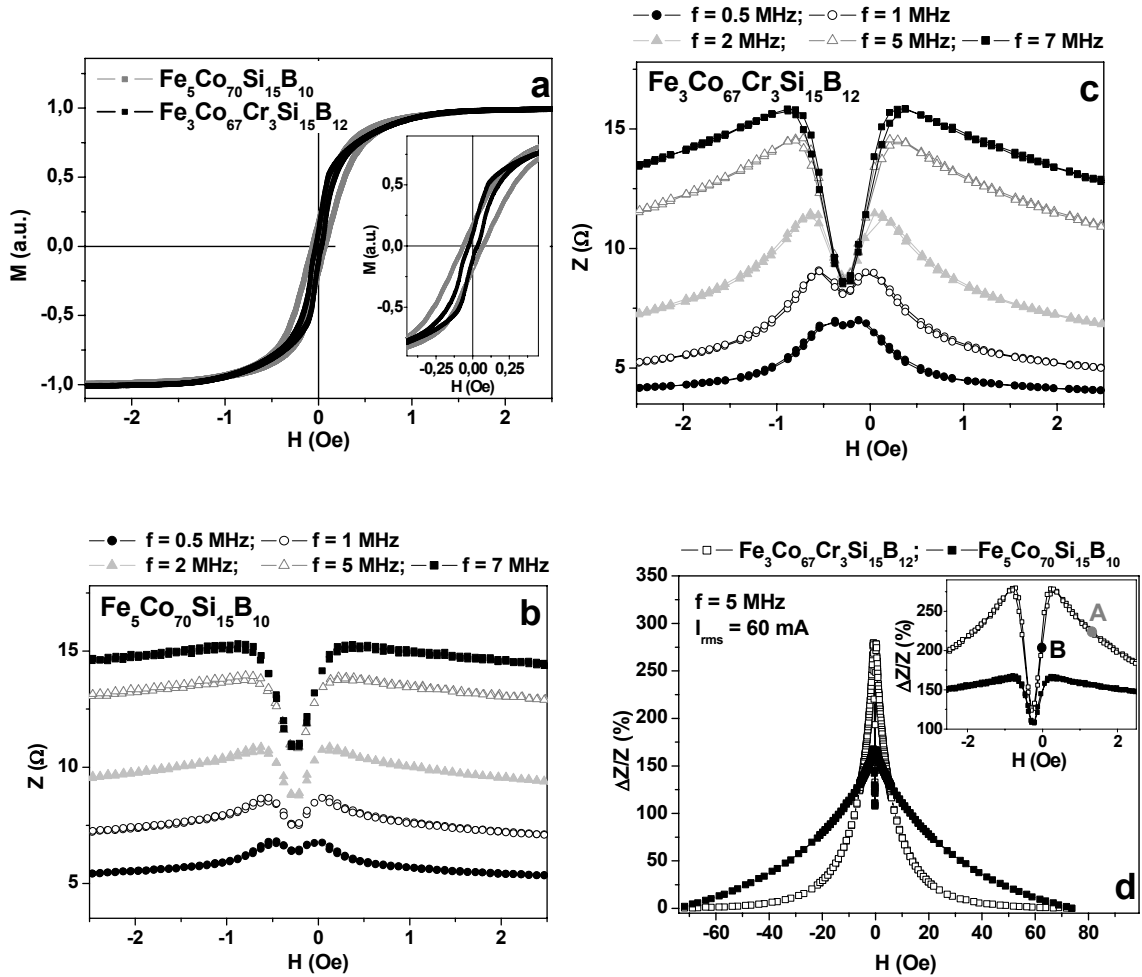

Figure 2.

Mass loss in a regime of passive exposure for different fluids (see also Table 2): a-d) - $Fe_5Co_{70}Si_{15}B_{10}$ – I and $Fe_3Co_{67}Cr_3Si_{15}B_{12}$ – I samples; d) – A and B are $Fe_5Co_{70}Si_{15}B_{10}$ –I samples separately subjected to the same treatment; inset shows x-ray diffraction of the as-quenched and $Fe_5Co_{70}Si_{15}B_{10}$ –II ribbon after passive exposure to urine during 30 hours.

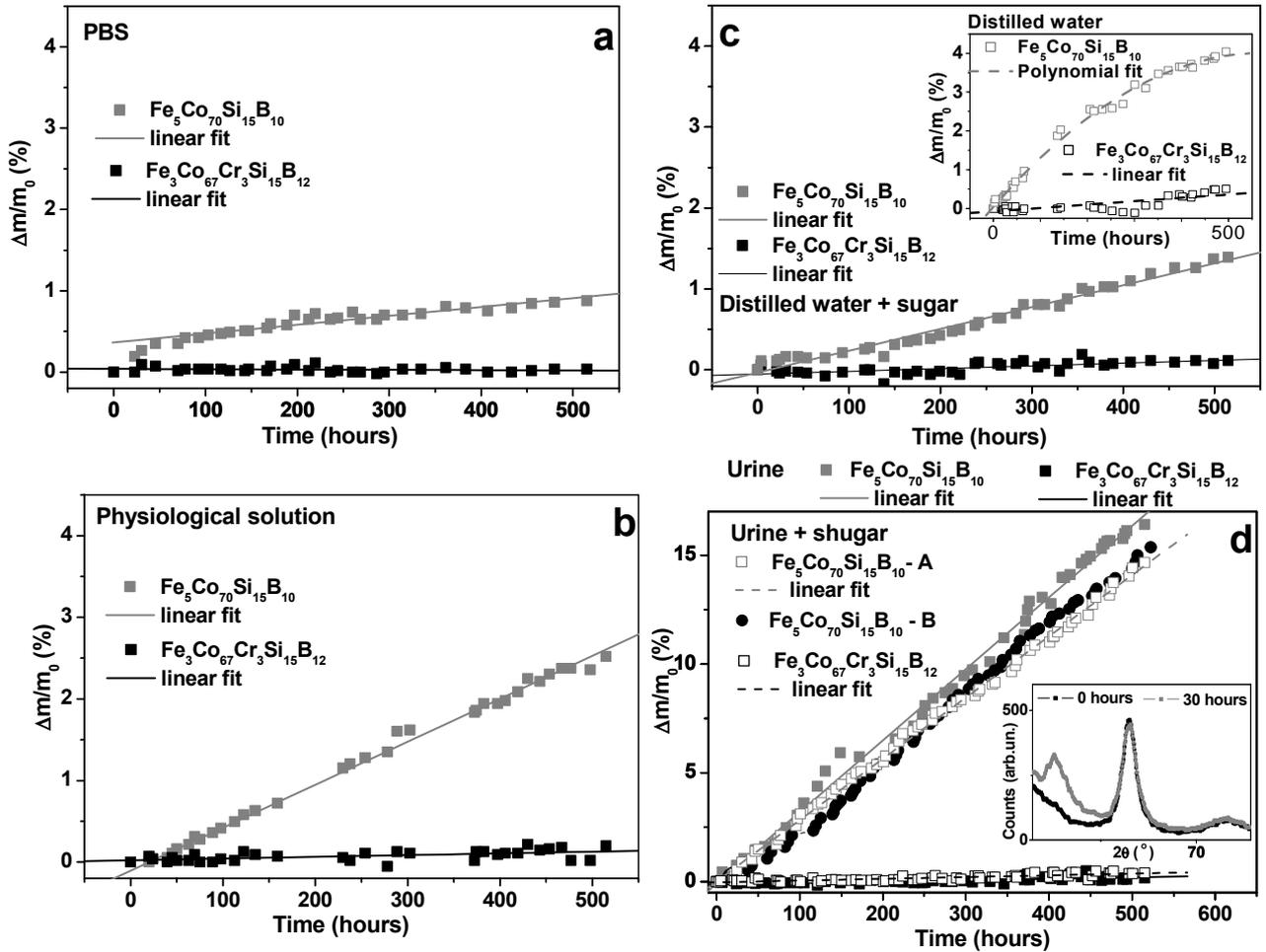

Figure 3

a) The dependence of the mass loss rate of the amorphous ribbons as a function of concentration of $H_3PO_4$ in passive regime.

b) The dependence of the mass loss of Co-based ribbons: A is the $Fe_5Co_{70}Si_{15}B_{10}$ − I sample; B is the $Fe_5Co_{70}Si_{15}B_{10}$ –II sample; C represents B data recalculated taking into account surface/volume ratio.

c) MI responses collected under regime of active surface modification for 7% $H_3PO_4$ as a function of the exposure time.

d) The dependence of the MI ratio maximum, $\Delta Z/Z_{max}$, and response in the field close to zero, $\Delta Z/Z_{min}$, as a function of time of treatment in 7% $H_3PO_4$ under regime of active surface modification.

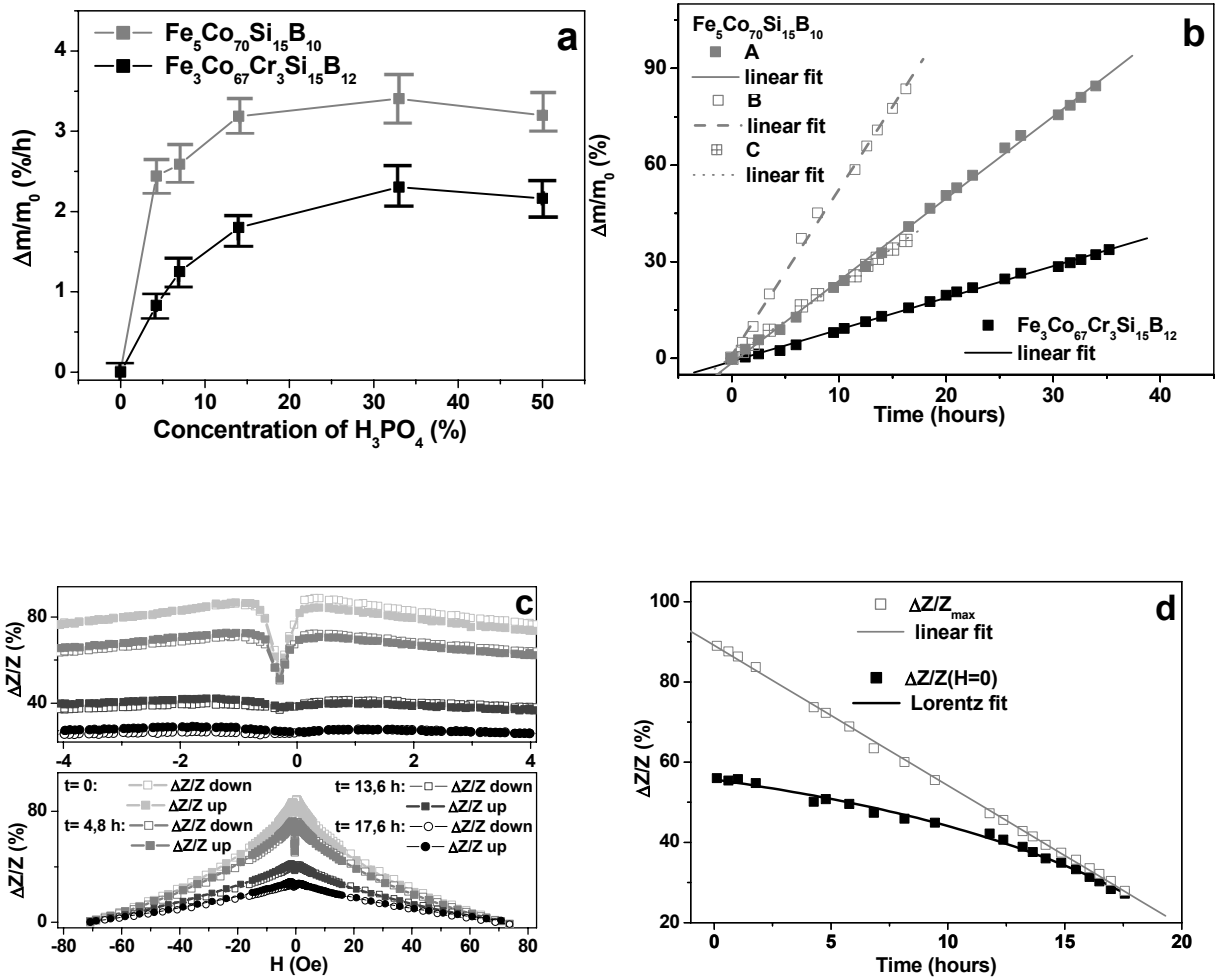

Figure 4

$Fe_5Co_{70}Si_{15}B_{10}$ ribbons: passive surface modification in urine:
a) MI responses of the amorphous ribbons for selected times of treatment.
b) Frequency dependence of the MI ratio maximum, $\Delta Z/Z_{max}$, and response in the field close to zero, $\Delta Z/Z_{min}$, for selected times of treatment.

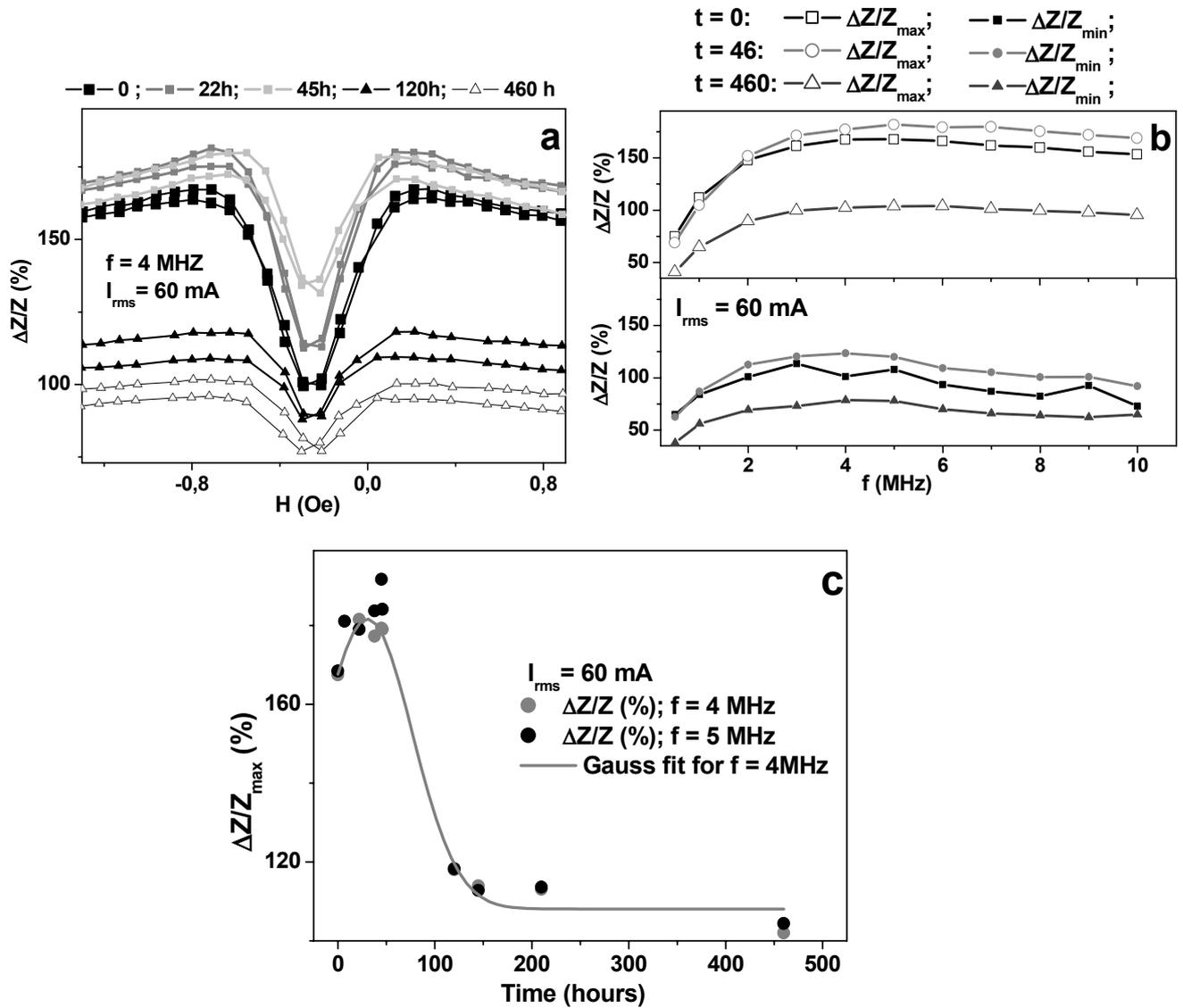

Figure 5.

Fe$_5$Co$_{70}$Si$_{15}$B$_{10}$ ribbons.
a-f) Passive surface modification in urine: a,c,e) – free side; b,d,f) – roller side.
g-h) Active surface modification in urine combined with MI measurements, both for free side: g) – bright field observation; h) – dark field observation.

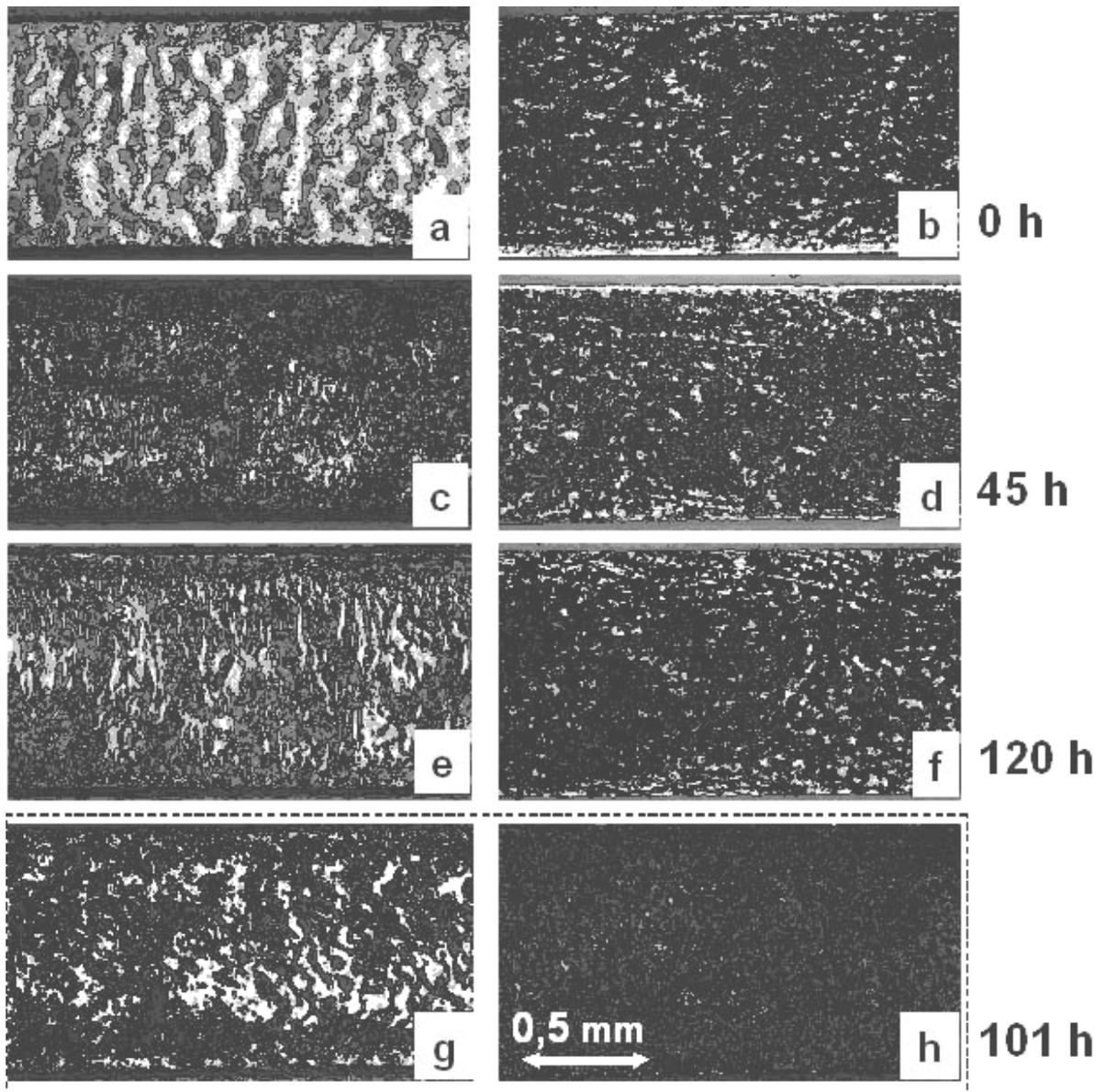

Figure 6.
$Fe_5Co_{70}Si_{15}B_{10}$ ribbons.
a-b) Active surface modification in urine combined with MI measurements.
Inset (a) : inductive hysteresis loops in as-quenched state and after treatment in a regime Y.

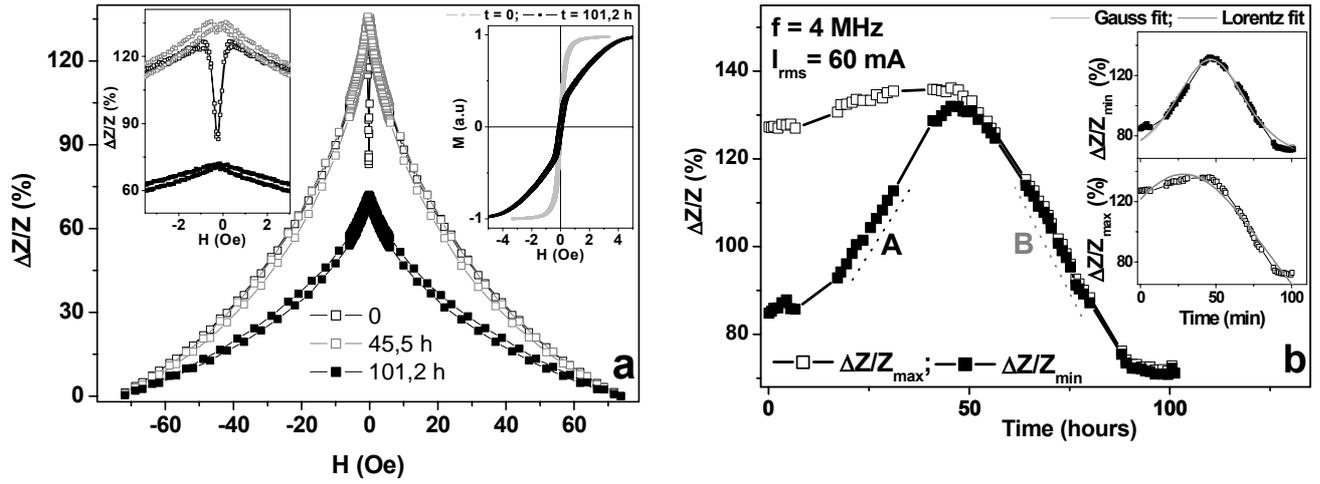